\documentclass{iau}
\usepackage{graphicx}
\title[Radiation Properties of NLS1s] 
{Radiation Properties of GeV Narrow Line Seyfert 1 Galaxies\footnote{Supported by the National Basic Research
Program (973 Programme) of China (Grant 2009CB824800), the National Natural Science Foundation of China (Grants
11078008, 11025313, 11133002, 10725313), Guangxi Science Foundation (2011GXNSFB018063).}}
\author[Jin Zhang et al.]   
{Jin Zhang $^1$, X. N. Sun$^2$, S. N. Zhang$^{3,1}$, E. W. Liang$^{2,1}$
 }
\affiliation{$^1$ National Astronomical Observatories, Chinese Academy of Sciences, Beijing, 100012, China
  \\ email: {\tt zhang.jin@hotmail.com} \\[\affilskip]
$^2$Department of Physics and GXU-NAOC Center for Astrophysics
and Space Sciences, Guangxi University, Nanning, 530004, China\\
$^3$Key Laboratory of Particle Astrophysics, Institute of High Energy Physics, Chinese Academy of Sciences,
Beijing, 100049, China}

\pubyear{2012}
\volume{290}  
\jname{Feeding compact objects: Accretion on all scales} \editors{C.M. Zhang, T. Belloni, M. M\'endez \& S.N.
Zhang, eds.}

\begin{document}
\maketitle
\begin{abstract}
The broadband SEDs of four gamma-ray NLS1s are compiled and explained with the leptonic model. It is found that
their characteristics and fitting parameters of the observed SEDs are more like FSRQs than BL Lacs.

\end{abstract}
\firstsection 
\section{Introduction}
Narrow-Line Seyfert 1 Galaxies (NLS1s) are a very peculiar class of active galactic neclei (AGN). A NLS1 is
characterized by an optical spectrum with narrow permitted lines FWHM (H$\beta$) $<$ 2000 km s$^{-1}$, the ratio
of [O III]$\lambda$5007 to H$\beta$ $<$ 3, and a bump due to Fe II (e.g., \cite[Pogge 2000]{Pogge2000}). NLS1s are
generally radio quiet, but a small fraction of them ($<7\%$) are radio loud (\cite[ Komossa et al.
2006]{Komossa2006}). Yuan et al. predicted (\cite[Yuan et al. 2008]{Yuan2008}) that the most radio loud NLS1s
should host relativistic jets, which dominate the high energy emission of NLS1s. So far, four NLS1s have been
detected as a new class of gamma-ray AGNs (\cite[Abdo et al. 2009a]{Abdo2009a}). Their GeV or TeV gamma-ray
emissions increase significantly our knowledge about their broadband SEDs, thus providing an opportunity to
constrain their model parameters and understand the analogies with and differences from those previously known
gamma-ray blazars.

\section{Model and Results}
\begin{figure}
\begin{center}
 \includegraphics[scale=0.7]{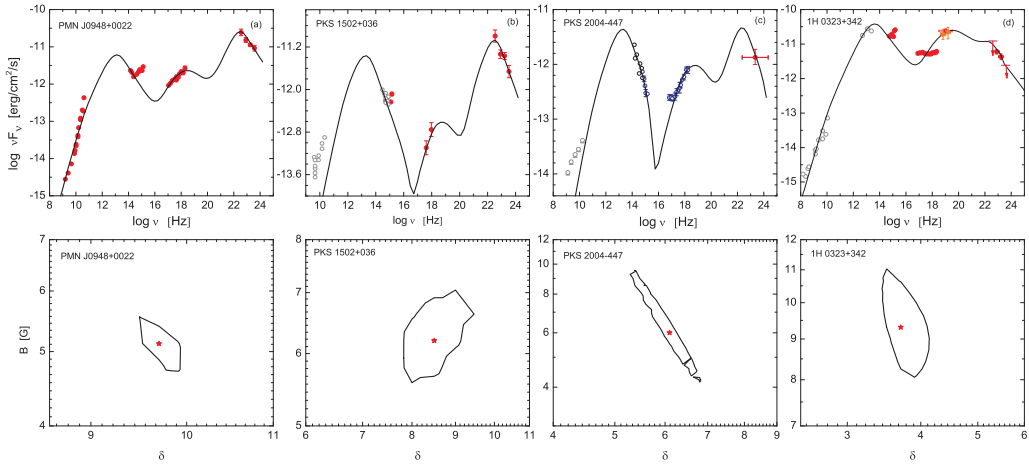}
 \caption{The observed SEDs ({\em scattered data points}) with model fitting ({\em lines}) for the four NLS1s and the contours in 1-$\sigma$ significance level in the $B$-$\delta$ plane. The observed data of PMN J0948+0022 are taken from \cite[Abdo et al. (2009b)]{Abdo2009b} and the data for other NLS1s are taken from \cite[Abdo et al. (2009a)]{Abdo2009a}. The {\em red solid circles} stand for the observation data with Fermi/LAT and simultaneous observation data in other bands. The {\em opened circles} indicate that they are not simultaneous with Fermi/LAT observations. The red stars in the contours stand for the best fitting parameter set of $\delta$ and $B$.}
   \label{fig1}
\end{center}
\end{figure}

The shapes of the observed SEDs of NLS1s seem like that of flat spectrum radio quasars (FSRQs), and the
contributions of inverse Compton (IC) by the external field photons from their broad line regions (BLRs) need to
be considered. The radiation from a BLR is assumed to be a blackbody spectrum and the corresponding energy
densities seen in the comoving frame are simply assumed as $U^{'}_{\rm BLR}=3.76\times10^{-2}\Gamma^{2}$ erg
cm$^{-3}$ (\cite[Ghisellini \& Tavecchio 2008]{Ghisellini2008}), where we take $\Gamma=\delta$. Following the
methods described in (\cite[Zhang et al. 2012]{Zhang2012}), the SEDs of NLS1s are fitted well by the
syn+SSC+IC/BLR model, where we do not consider the thermal emission of accretion disk in the optical-UV band and
the possible corona emission at X-ray energies. The fitting results are shown in Fig.\,\ref{fig1}.

The contours of the probability $p$ ($p\propto e^{-\chi^2}$) for the parameters $B$ and $\delta$ in 1 $\sigma$
significance level are also presented in Fig.\,\ref{fig1}. It is found that the tighter constraints on $B$ and
$\delta$ are obtained with the better observed SEDs. The range of $\delta$ for these NLS1s is from 3 to 10 and
the magnetic field strengths are from 4 G to 11 G. The values of $\gamma_{\rm b}$ are several hundreds. It is
clear that the physical parameters of NLS1s are more like that of FSRQs than BL Lacs.

\section{Summary}
The observed broadband SEDs of the four gamma-ray NLS1s, which are dominated by jet emission, can be well
explained by the leptonic jet model. The shapes and fitting parameters of the observed SEDs of these NLS1s are
more like that of FSRQs than BL Lacs. FSRQs, BL Lacs and, now, RL-NLS1s are characterized by the presence of a
relativistic jet. The differences in their observed SEDs and the fitting parameters may be mainly determined by
their different masses and accretion rates. The high accretion rates, which is close to the Eddington rate, and
low black hole masses of NLS1s may imply they are in an early phase of blazar evolution. A detailed analysis on
the comparison of physical properties for the three GeV-TeV AGNs, NLS1s, FSRQs and BL Las, will be presented in
details elsewhere.


\begin{thebibliography}{}

\bibitem[Abdo et al. (2009a)]{Abdo2009a} {Abdo, A.A., Ackermann, M., Ajello, M., et al.} 2009a, \textit{ApJ}, 707, L142
\bibitem[Abdo et al. (2009b)]{Abdo2009b} {Abdo, A.A., Ackermann, M., Ajello, M., et al.} 2009b, \textit{ApJ}, 707, 727
\bibitem[Ghisellini \& Tavecchio (2008)]{Ghisellini2008} {Ghisellini, G., \& Tavecchio, F.} 2008, \textit{MNRAS}, 387, 1669
\bibitem[Komossa et al. (2006)]{Komossa2006} {Komossa, S., Voges, W., Xu, D., et al.} 2006, \textit{AJ}, 132, 531
\bibitem[Pogge (2000)]{Pogge2000} {Pogge, R.W.} 2000, \textit{Nature}, 44, 381
\bibitem[Yuan et al. (2008)]{Yuan2008} Yuan, W., Zhou, H. Y.,  Komossa, S., et al. 2008, \textit{ApJ}, 685, 801
\bibitem[Zhang et al. (2012)]{Zhang2012} {Zhang, J., Liang, E.-W., Zhang, S.-N., \& Bai, J.~M.} 2012, \textit{ApJ}, 752, 157

\end{thebibliography}
\end{document}